\documentclass[twocolumn,nofootinbib,amsmath,prl,aps,superscriptaddress,tightenlines,preprintnumbers]{revtex4}

\usepackage{graphicx}
\usepackage{braket}
\usepackage[usenames]{color}

\usepackage{mathrsfs}

\newcommand{\Ga}{\Gamma}

\newcommand{\nn}{\nonumber}
\newcommand{\ov}{\overline}

\newcommand{\cB}{\mathcal B}

\newcommand{\cA}{\mathcal A}

\newcommand{\mev}{{\rm MeV}}
\newcommand{\m}{{\ov m}}

\newcommand{\vareps}{ \varepsilon}

\def\nn{\nonumber\\ }
\def\rd{{\rm d}}
\def\abs#1{\left| #1 \right| }

\def\ci{c_1}
\def\cii{c_2}
\def\ciii{c_3}

\def\clr{X_\ell}
\def\cdiff{Y_{\ell}}
\def\gv{ g_V }
\def\fstate{\mathcal{F}}

\begin{document}

\newcount\hour \newcount\minute
\hour=\time \divide \hour by 60
\minute=\time
\count99=\hour \multiply \count99 by -60 \advance \minute by \count99
\newcommand{\mydate}{\ \today \ - \number\hour :00}

\title{Probing the nature of the Higgs-like Boson via $h\to V \fstate$ decays.}

\author{
Gino Isidori${}^{1,2}$, Aneesh V. Manohar${}^3$ and Michael Trott${}^2$\\
${}^1$ INFN, Laboratori Nazionali di Frascati, I-00044 Frascati, Italy \\
${}^2$ Theory Division, Physics Department, CERN, CH-1211 Geneva 23, Switzerland \\
${}^3$ Department of Physics, University of California at San Diego, La Jolla, CA 92093}

\begin{abstract}
We give a general decomposition of the $h \rightarrow V \fstate$  amplitude where $V=\{W^\pm,Z^0\}$ and $\fstate$ is a generic  leptonic  or hadronic final state,
in the standard model (SM), and in the context of a general effective field theory. 
The
differential distributions for $\fstate=\ell^+\ell^-, \ell \overline \nu$  ($\ell =e,\mu$) are reported, and we show how such distributions can be used to determine modified Higgs couplings that cannot be directly extracted from a global fit to Higgs signal strengths. We also demonstrate how rare 
$h \rightarrow V  P$ decays, where $P$ is a pseudo-scalar meson, with SM rates in the $10^{-5}$ range, 
 can  be used to provide complementary information on the couplings of the newly discovered Higgs-like scalar and are an interesting probe of the vacuum structure of the theory.
\end{abstract}

\maketitle
\newpage

\paragraph{\bf I. Introduction:}
Characterizing the properties of the newly discovered scalar boson at the LHC~\cite{Higgs}  is of central importance to determine experimentally
the nature of electroweak symmetry breaking, and to investigate the possibility of physics beyond the Standard Model (SM).
It is particularly important  to determine how the new boson couples to the $SU(2)_L\times U(1)_Y$ gauge fields, since these couplings are directly related to symmetry breaking and gauge boson mass generation.
The 125~GeV boson cannot decay into an on-shell pair of massive gauge bosons, but it can decay via $h \to V V^*$, $V^* \to \fstate$, where one of the bosons is off-shell.

The purpose of this paper is to show that the offshellness of $V^\star$ can be viewed as a virtue in Higgs studies. It allows one to measure decays to final states $\fstate$ that would not be accessible if on-shell decays were allowed, and the additional decay channels  increase the sensitivity to new-physics (NP) effects, as they affect kinematic distributions of $\fstate$ in addition to the total rate. We demonstrate this conclusion using two examples: (a) $\fstate$ is a pair of light leptons  $\ell^+\ell^-$ or $\ell\nu$,  with $\ell =e,\mu$, and  (b) $\fstate=P$ is a hadronic state composed of a single pseudoscalar or vector meson.
 
In the two-lepton case, most of the interesting information is encoded in the two-dimensional 
kinematical distributions of the leptons in the Higgs rest frame. We analyze such distributions both in the SM, and in
the context of a general effective field theory (EFT) approach, neglecting lepton masses. We show that these distributions, which will soon be accessible at the LHC with increasing 
statistics, contain information that cannot be directly extracted from a global fit to signal strengths.
In the $h \rightarrow  VP$  case we show how these rare processes, with SM rates in the $10^{-5}$ range, 
 can provide a complementary tool to extract Higgs properties not easily accessible from the purely leptonic modes. 

\paragraph{\bf II. Amplitude Decomposition:}
Consider the  $h\to V \fstate$ amplitude where $V=\{W^\pm,Z^0\}$ is an on-shell massive weak gauge boson 
while $\fstate$ is a final state generated at  tree level by the electroweak charged or neutral currents,  
\begin{align}
\label{currentdefn}
\mathcal{L}_{J} = \frac{e }{ \sqrt{2}\, \sin \theta_W}  J_\mu^{\pm} W^\mu_{\pm}   + \frac{e}{\sin \theta_W \cos \theta_W}  J_\mu^0 Z^\mu \, .
\end{align}
Let $J^{\fstate,V}_\mu =\braket{ \fstate | J^V_\mu |0 }$ be the 
matrix element relevant for $V \to \fstate$. The decay amplitude
$\cA\!\left[ h \to V(\vareps,p) \fstate(q)\right ] $  can be decomposed in terms of four independent Lorentz structures,
which we define as
\begin{align}
 \cA^{\fstate}_V &=  C_V \gv^2 m_V  \frac{\vareps_\mu  J^{\fstate}_\nu}{(q^2 - m_V^2)}  \left[f^V_1({q}^2) g^{\mu\nu}  
 +  f^V_2({q}^2)  {q}^\mu {q}^\nu \right.    \label{eq:gendec} \\
 &\,  \left. +  f^V_3({q}^2)({p}\cdot {q}~g^{\mu\nu} -{q}^\mu {p}^\nu)  + f^V_4({q}^2) \epsilon^{\mu\nu\rho\sigma} {p}_\rho {q}_\sigma \right] \nonumber.
\end{align}
Here  $\gv=\{g_2,g_2/\cos\theta_W\}$, $g_2=e/\sin\theta_W$, $C_V = \{1/\sqrt{2},1\}$ 
are the coupling and normalization factors for $V=\{W^\pm,Z^0\}$. 
Throughout this paper $p$ will denote the $V$ four-momentum and $q$ the total four-momentum of the final state $\fstate$.  In writing Eq.~(\ref{eq:gendec}), we have used $p \cdot \vareps=0$ for physical $V$ bosons but we have not made any assumption about the angular momentum of the $\fstate$ state.
 We will also use the  dimensionless variables $\rho = m_V^2/m_h^2$,  $\hat q^\mu = q^\mu/m_h$, $\hat p^\mu = p^\mu/m_h$.
$f_i^Z$ are real, and $f_i^{W^\pm}$ are complex conjugates of each other. 
For a $0^+$ scalar $h$, $\text{Im} f_{1,2,3}^W\not = 0$ 
and $\text{Re} f^{W,Z}_4 \not =0$ violate $CP$.

In general, the $f_i(q^2)$ are four independent dimensionless form factors. At $q^2=m_V^2$,  two of them satisfy the relation 
\begin{align}
f^{V}_{1}(m_V^2)  =  - m_V^2 \, f^{V}_{2}(m_V^2)\,,
\label{pole}
\end{align}
dictated by the requirement that the pole of the amplitude when $q^2 \to m_V^2$ corresponds 
to the exchange of an on-shell $V$. In the SM, $f_1^{\rm SM}= 1$, $f^{\rm SM}_{2} = -1/m_V^2$, and $f_{3,4}^{\rm SM}=0$.

The different form-factors can be probed by using different final states. The differential decay rate summing over $V$ polarizations is
\begin{align}
\rd\Gamma = \frac{\pi^2  C_V^2  \gv^4 m_V^2}{2 m_h}\   \frac{ {\mathcal M}^{\mu \nu}  J^{\fstate}_{\mu}  J_{\nu}^{\fstate \dagger }}{ ({q}^2 
-m_V^2)^2 }\  \lambda(\hat{q}^2,\rho) \    \rd {q}^2\, \rd\Phi_{\fstate}\,,
\label{eq:dGgen}
\end{align}
where $ \lambda(\hat{q}^2,\rho) = \sqrt{(1 + \hat{q}^2 - \rho)^2 - 4 \hat{q}^2}$ and
\begin{align}
\rd\Phi_{\fstate} =  \prod_{i=1\ldots n_{\fstate}}  \frac{ d^3 k_i}{2E_i (2\pi)^3} \delta^4\left(q- \sum k_i \right)\,,
\end{align}
denotes the phase space of the final state $\fstate$. 
In ge\-ne\-ral, the tensor structure  ${\mathcal M}^{\mu \nu}$ depends on the form-factors $f_i^V({q}^2)$, but
it is simplified for specific choices of $\fstate$.
For the two-lepton final state the currrent is conserved, 
${q}_\mu J^{\fstate}_{\mu}=0$, when the lepton masses are neglected, and ${\mathcal M}^{\mu \nu}$  becomes
\begin{align}
 \mathcal{M}^{\mu \, \nu}_{\ell \ell} &=  - g^{\mu  \nu}  \left( \abs{  h_1}^2 - \abs{f_4}^2\left[{p}^2 {q}^2-({p} \cdot {q})^2\right] \right)  \nn
& + {p}^\mu  {p}^{\nu} \left(\frac{\abs{f_1}^2}{m_V^2} - {q}^2  \abs{f_3}^2 - {q}^2  \abs{f_4}^2 \right)\nn
&+2i\, \text{Im} (h_1^* f_4)\ \epsilon_{\mu \nu \rho \sigma  }p^\rho q^\sigma\,,
\label{eq:Fcons}
\end{align}
where $h_1=f_1 + (p \cdot q) f_3$.
In the case of a single pseudoscalar meson, the current assumes the form $J^\fstate_{\mu}\propto q_\mu$ and 
\begin{align}
  \mathcal{M}^{\mu \, \nu}_{P} =  \left( - g^{\mu  \nu}  +\frac{{p}^\mu {p}^\nu}{m_V^2}  \right) 
  \abs{ f_1+ {q}^2 f_2 }^2\,.
  \label{eq:FP}
\end{align}
$f_{3,4}$ are present only in  $\mathcal{M}^{\mu \, \nu}_{\ell \ell}$, while $f_{2}$ is present only in $ \mathcal{M}^{\mu \, \nu}_{ P}$.
Some information on the $h$ decay amplitude is lost when summing over $V$ polarizations.
This information is essential to determine the spin and parity  of  $h$  (see e.g.~Ref.~\cite{spin})
but is  less relevant once we assume $h$ to be a $0^+$ state, as in most realistic NP models. 

{\bf III. {\bf Modifications of the Spectra:}}
In order to investigate how the $f_i(q^2)$ can vary in possible extensions of the SM, we consider a
general EFT that contains the SM scalar sector in the low energy theory. The EFT includes explicitly the Goldstone bosons associated with the breaking of $ SU(2) \times U(1)_Y \rightarrow U(1)_Q$, as well as the remaining SM field content
 with a nonlinear realization of the $ SU(2) \times U(1)_Y$ symmetry and a singlet scalar field 
 $h$~\cite{Bagger:1993zf,Koulovassilopoulos:1993pw,Burgess:1999ha,Grinstein:2007iv,Contino:2010mh}.
 The Goldstone bosons eaten by the $
W^\pm, Z$ bosons are denoted by $\pi^a$ where $a = 1,2,3$, and are grouped as $ \Sigma(x) = \exp [{i \tau^a \, \pi^a/v}]$.
The operators that are of interest in this work are
\begin{align}
O_{LO} &= \frac{v  \, \ci}{2}\  h\, {\rm Tr} \left[ (D_\mu \Sigma)^\dagger \, D^\mu \Sigma\right],  \nn
\mathcal{O}_W &=  \frac{g_2  \, \cii}{ v}\ h \, D_\mu W_a^{\mu  \nu} {\rm Tr} \left[\Sigma^\dagger \, i \tau^a  \overleftrightarrow{D}_\nu \Sigma\right],  \nn
\mathcal{O}_{W\! \partial H} &= \frac{g_2  \, \, \ciii  }{ v}\ (\partial_\nu h) \, W_a^{\mu \nu}  {\rm Tr} \left[ \Sigma^\dagger \, i \tau^a \,  \overleftrightarrow{D}_\mu \Sigma \right]\,. \label{eq:Op}
\end{align}
Here $c_i$ are unknown Wilson coefficients, with $c_i^{\rm SM} = (1,0,0)$. The complete operator basis to sub-leading order for theories of this form is given in Ref.~\cite{Alonso:2012px,Buchalla:2013rka}.

The subleading operators in Eq.~(\ref{eq:Op}) are chosen as they do not violate custodial symmetry at $g_1=0$; this simplifying choice is made to demonstrate the utility of the differential spectra. 
For simplicity we have also neglected possible NP effects giving rise to local operators coupling $h$, $V$ and the 
fermonic currents directly, or modifying the currents themselves. These effects could still be described by the general decomposition 
in Eq.~(\ref{eq:gendec}), but with contributions to the amplitude that will not have a pole as $q^2 \to m_V^2$, and can contain
non-universal  ($\fstate$-depedent)  form factors.

We find that the projection of the custodial symmetry preserving operator basis  to the form factor basis is 
\begin{align}
f_1^V({q}^2)  &=   \ci   +  g_2^2  \, (\cii \, +\ciii) \, \left(1 + \frac{{q}^2}{m_V^2} \right)~, \nn
f_2^V({q}^2)  &= - \frac{1}{m_V^2} \, \left[\ci  + 2 \, g_2^2  \, (\cii \, + \ciii) \right]~,  \nn
f_3^V({q}^2) &= \frac{2 \, g_2^2 }{m_V^2}\ciii~,  \quad \quad f_4^V({q}^2) = 0~. 
\end{align}
The three parity-conserving form factors correspond to three independent combinations of the 
parameters of the EFT Lagrangian, and only one combination is determined by the total decay rate.
The dependence of the differential rate on  $f_i({q}^2)$ in Eqs.~(\ref{eq:Fcons})--(\ref{eq:FP})
offers the opportunity to determine the individual form-factors, and hence the independent operator coefficients with sufficient data.

{\bf IV. {\bf  $ \fstate =\ell^+\ell^-,\ell\nu$:}}\label{spectra}
There are two kinematic variables needed to describe the final state, after averaging over lepton spins.
Two convenient choices are either the two lepton energies in the $h$ rest frame $(E_{1,2}$),
or the lepton invariant mass ${q}^2$ and $c_\theta\equiv\cos\theta$, where $\theta$ is 
the angle between the lepton axis in the dilepton rest frame and the Higgs momentum.
For these two cases we can write ($y_i=2E_i/m_h$, with $i=1$ for the lepton and $i=2$ for the antilepton)
\begin{align}
\frac{ \rd^2 \Gamma}{\rd  y_1  \rd y_2} = \frac{ 2 m_h^2}{ \lambda(\hat{q}^2,\rho)  } \frac{ \rd^2 \Gamma}{\rd  q^2  \rd c_\theta }
= \frac{C_V^2 \gv^4 m_V^2 m_h}{256 \pi^3}  \frac{ \left[{\mathcal M}^{\mu \nu}_{\ell \ell}  J^{\fstate}_{\mu}  J_{\nu}^{\fstate\dagger }\right]}{( q^2 -m_V^2)^2} ~,
\end{align}
Neglecting lepton masses, the term between square brackets can be evaluated from Eq.~(\ref{eq:Fcons}) using 
\begin{align}
&\frac{1}{\clr m_h^2} \sum_{\rm \ell \,spins} \, J \cdot J^\dagger = - 2 \hat{q}^2 = 2 (1-\rho -y_1 -y_2)  \nn
&\frac{1}{\clr m_h^4}\sum_{\rm \ell \,spins} ({p} \cdot J) ({p} \cdot J^\dagger) =-\hat q^2+y_1 y_2\nn
&\hspace{3.8cm} =\frac14 \lambda^2(\hat q^2,\rho) \left(1-c_\theta^2\right)\nn
& \frac{1}{\cdiff m_h^4}\sum_{\rm \ell \,spins} \epsilon_{\mu \nu \rho\sigma}J^{\mu}  J^{\nu \dagger} p^\rho q^\sigma = i  \hat q^2 (y_1-y_2)
 \label{eq:spin}
\end{align}
where $\clr = (g^\ell_R)^2 + (g^\ell_L)^2$, $\cdiff=(g^\ell_R)^2 - (g^\ell_L)^2$.

The general expression of the double differential energy distribution 
can be deduced from Eqs.~(\ref{eq:dGgen})--(\ref{eq:spin}).
\begin{widetext}
\begin{align}
&\frac{ \rd^2 \Gamma }{\rd  q^2 \rd c_\theta  } = \frac{ C_V^2 \gv^4 m_V^2 }{256 \pi^3 m_h} 
\frac{\lambda(\hat{q}^2,\rho)}{\left(q^2-m_V^2\right)^2}\biggl\{
 \clr  q^2 \left[\abs{f_1 + \frac12 \left(m_h^2-q^2-m_V^2\right) f_3
}^2+\frac14 m_h^4 \lambda^2(\hat{q}^2,\rho)\abs{f_4}^2\right]\nn
&+  \frac18 \clr m_h^4 \lambda^2(\hat{q}^2,\rho) \left[\frac{\abs{f_1}^2}{m_V^2} - {q}^2  \abs{f_3}^2 - {q}^2  \abs{f_4}^2 \right]
\left(1-c_\theta^2\right)-\,\cdiff\,  \text{Im} \left[\left(f_1^*+\frac12(m_h^2-q^2-m_V^2)f^*_3\right) f_4\right] m_h^2 q^2  \lambda(\hat{q}^2,\rho) c_\theta\biggr\}
\end{align}
\end{widetext}
with $0 \leq \hat q^2 \leq (1-\sqrt{\rho})^2$ and $-1 \leq c_\theta \leq 1$. The $\hat{q}^2$ spectrum is particularly simple and sensitive  to possible deviations from SM.

In the SM, 
$f_1 = 1$, $f_{3,4} =0$, and the double differential rate is
\begin{align}
&\frac{1}{\mathcal{N_{\rm SM}}} \frac{ \rd^2 \Gamma^{\rm SM}}{\rd  y_1 \rd y_2  } =
  \frac{\rho^2 + \rho(y_1 + y_2 -\frac{3}{2}) + \frac{1}{2}(1-y_1)(1-y_2)}{(1-y_1-y_2)^2}
\end{align}
where $\mathcal{N_{\rm SM}}=\clr  C_V^2 \gv^4  m_h/(128 \pi^3) $
and $0 \leq y_1 \leq (1-\rho)$ and $ (1-\rho - y_1) \leq y_2 \leq (1-\rho-y_1)/(1-y_1)$. The lepton energy spectrum is
\begin{align}
 \frac{2}{\mathcal{N_{\rm SM}}} 
\frac{\rd \Gamma^{\rm SM} }{\rd  y_1} &= \frac{y_1 (1-y_1 - \rho) [2 \rho^2 - \rho + y_1(1 -y_1))}{\rho(\rho - y_1(1-y_1)]} \nn
&   + (2 \rho + y_1 - 1) \log \left[\frac{\rho - y_1 (1- y_1)}{\rho (1-y_1)} \right].
\label{eqn:dist3}
\end{align}

\begin{figure*}[t]
  \centering
  \includegraphics[width=7.9cm]{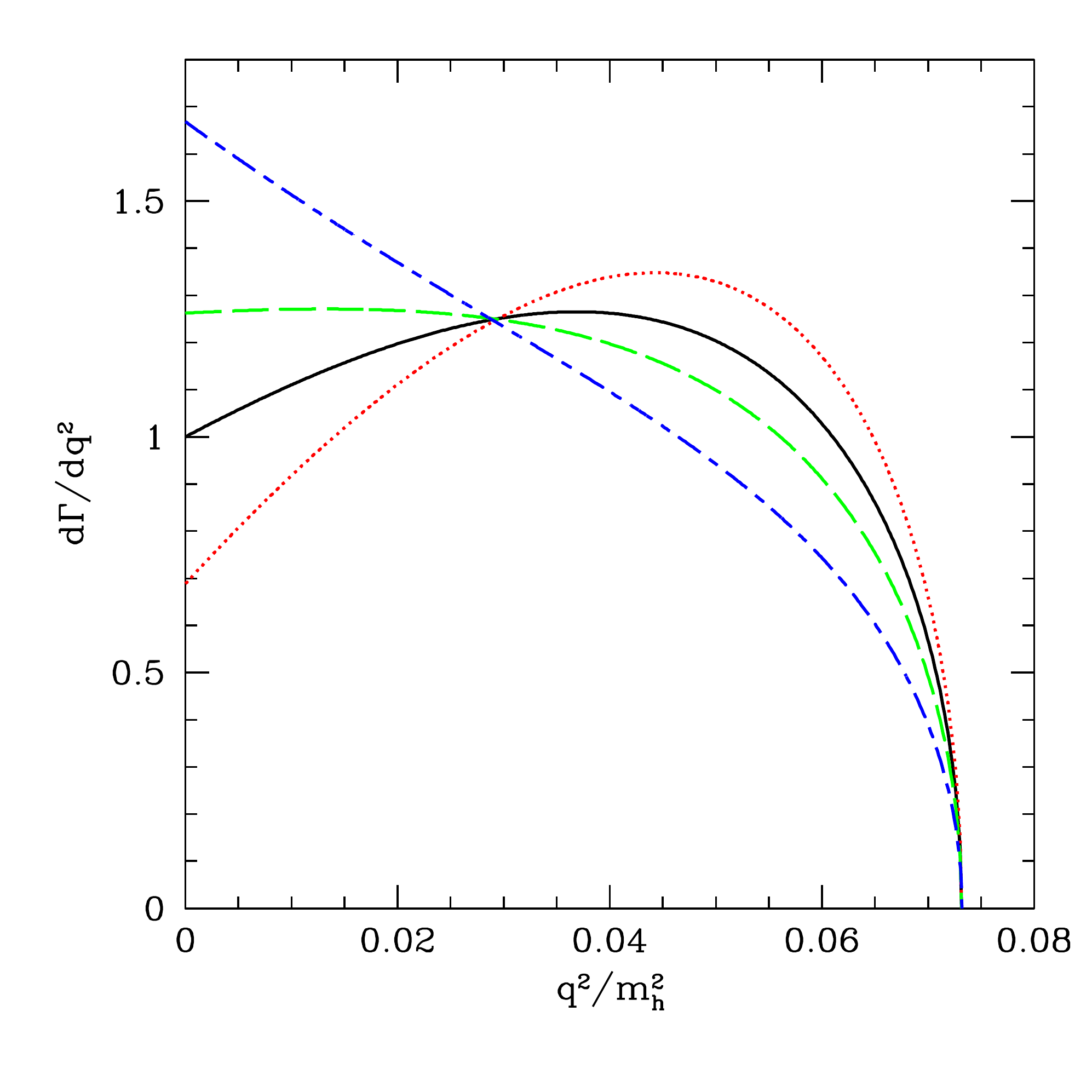}  \includegraphics[width=7.9cm]{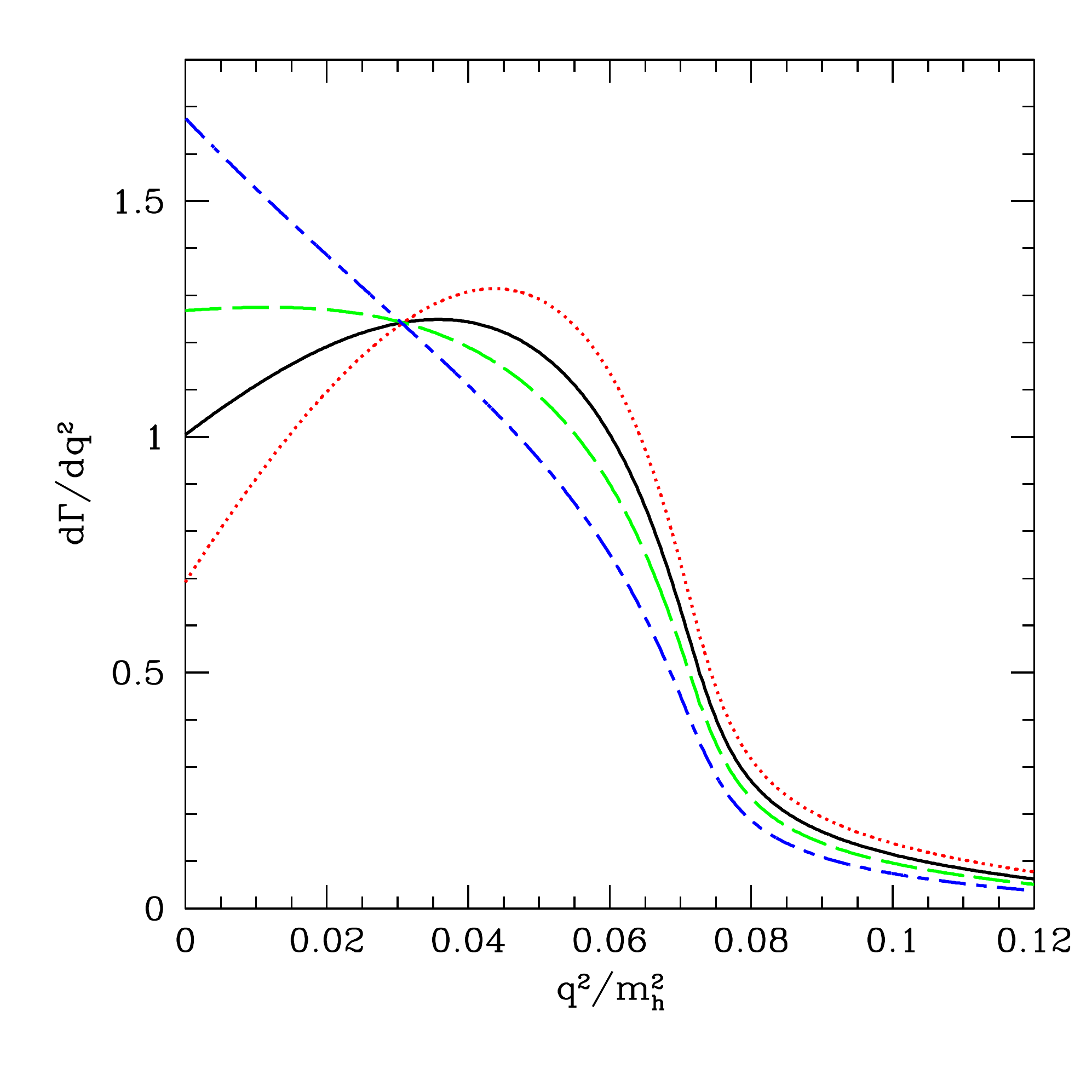}
  \vskip -1 true cm
 \includegraphics[width=7.9cm]{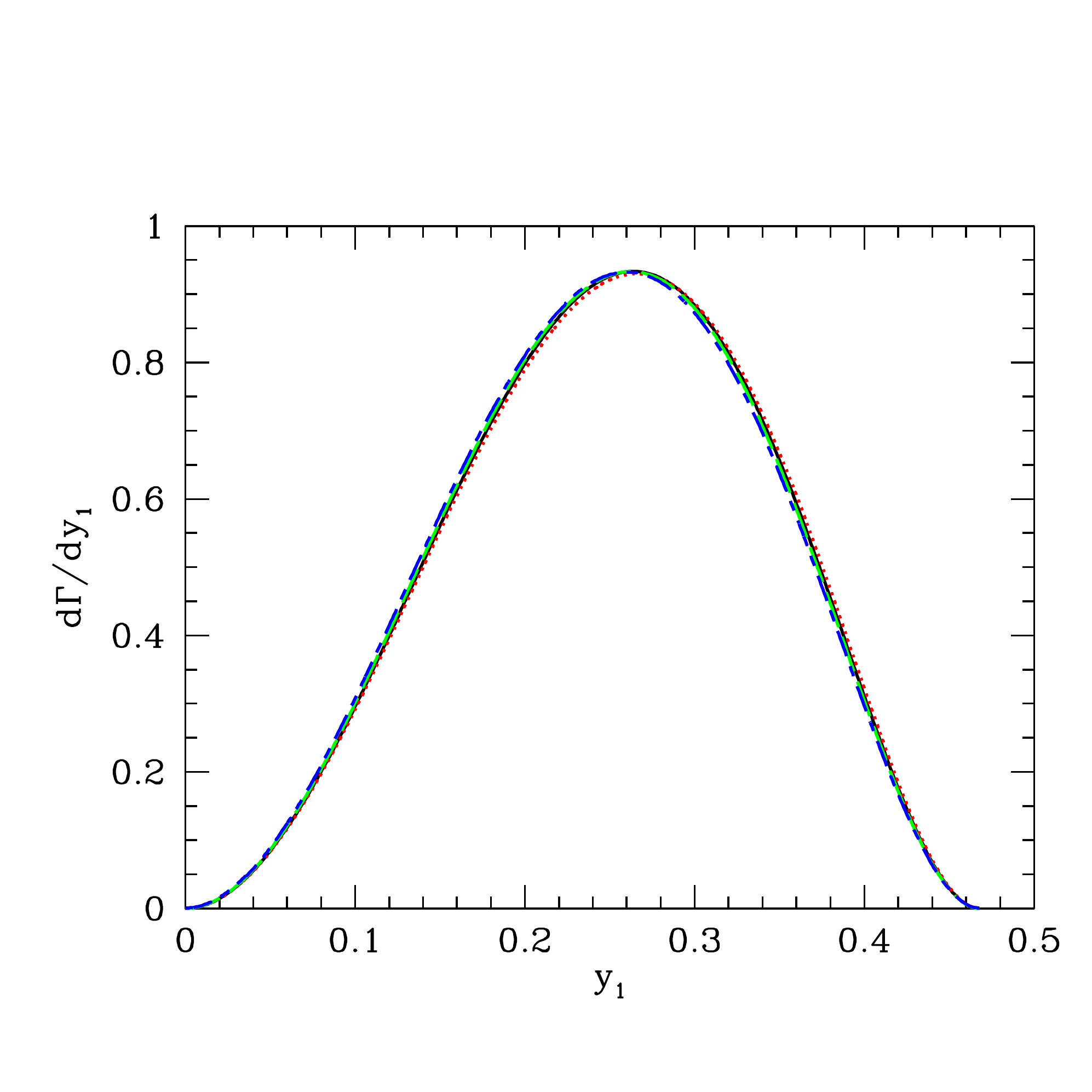}  \includegraphics[width=7.9cm]{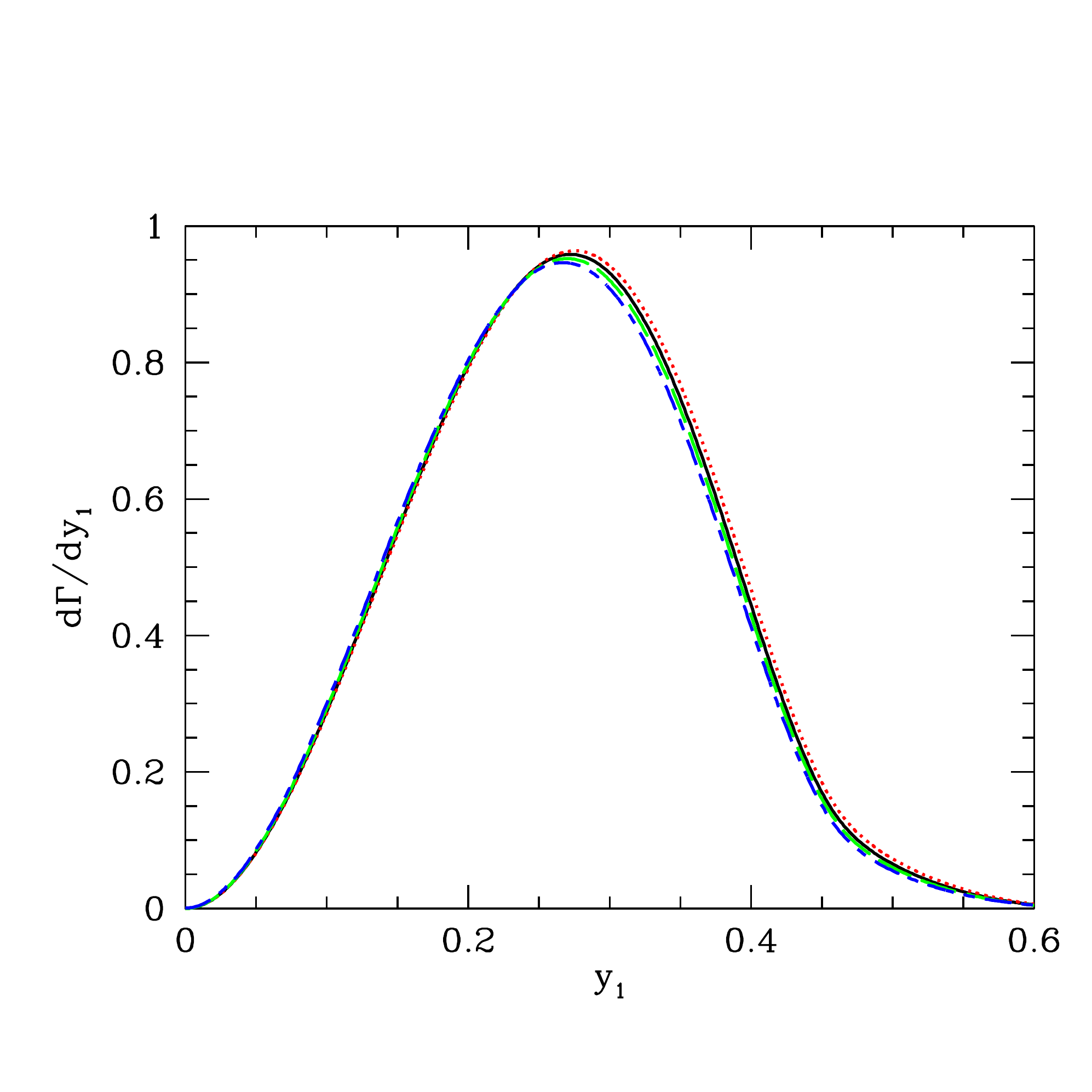}
  \caption{$\rd \Gamma/\rd {q}^2$ and $\rd\Gamma/\rd y_1$ spectra (arbitrary units) for different values of  EFT parameters 
  chosen to leave the total rate unchanged. The plots are for $V=Z$, and 
  in both plots the black (full) curve corresponds to the SM, $c_i=(1,0,0)$, the red (dotted) curve is  for $c_i=(0.82,-0.8,0.8)$, 
  the green (dashed) curve for $c_i=(1.18,0.5,-0.6)$, and the blue (dash-dotted)  curve for $c_i=( 1.30,  1.5,  -1.5)$. The curves in the upper plot accidentally pass through a common point, for our choice of parameters. The values of $c_i$ have been chosen for illustrative purposes, and are a bit larger than expected from EFT power counting. NDA \cite{Manohar:1983md} indicates that $c_{2,3} \sim \mathcal{O}(v^2/\Lambda^2)$ and $c_1 - 1 \sim \mathcal{O}(v^2/\Lambda^2)$. The left plots are for $m_V=m_Z$, and the right plots are convoluted with a Breit-Wigner distribution of width $\Gamma_Z$ over the mass range $m_Z \pm 10$\,GeV.}
\label{fig:spectra}
\end{figure*}
The usefulness of these differential distribution is illustrated in Fig.~\ref{fig:spectra}, where we compare different 
spectra, with $c_i$ chosen to leave the total rate unchanged.\footnote{
A detailed analysis of the present experimental constraints on the $c_i$ 
is beyond the scope of this work. However, we note that  in an EFT with a linear realization of $SU(2)_L \times U(1)_Y$ in the scalar sector, it has been shown that current experimental bounds on the Wilson coefficients still allow a variation of the spectra on the order of $\sim 20 \%$ \cite{carrell}, similar to the variation shown in Fig.~\ref{fig:spectra}. For a nonlinear realization of $SU(2)_L \times U(1)_Y$, there are no direct bounds on the Wilson coefficients from processes without a Higgs-like boson. }
 The $\rd \Gamma/\rd {q}^2$ spectrum exhibits large shape variations, which can be  directly probed experimentally, and used to 
constrain the EFT.
The  variation in the $q^2$ spectrum is due to a modified weighting of the terms in Eq.~(\ref{eq:spin}), which have a different $q^2$ dependence.

On the other hand, the dependence of the lepton energy spectrum $\rd \Gamma/\rd y_1$ on $c_i$ is much weaker. Integrating over $y_2$
averages
over a wide range of $\hat{q}^2$. As a result the shape of the $d \Gamma/d y_1$ distribution is almost universal, even in presence of the NP effects in the 
EFT we consider. This  stability of $\rd \Gamma/\rd y_1$ does not make it uninteresting --- it provides a check for reconstruction errors or unaccounted for experimental systematics.
The area under the curve of this spectrum is the total decay rate, 
and deviates from the SM value while maintaining this common shape.

We have examined the possibility of using moments of the lepton energy and $q^2$ spectra to extract the Wilson coefficients of the
operators in the EFT. However, these moments depend weakly on the $c_i$.
A more promising observable is the asymmetry $\mathscr{A}$ (either differential or integrated)  given by weighting $\rd^2 \Gamma/\rd y_1\rd y_2$ by ${\rm sign} ( y_2 -\bar y_2)$
where $\bar y_2 = (1-\rho - y_1) (1-y_1/2)/(1-y_1)$ is the midpoint of the $y_2$ integration range. 
$\mathscr{A}$ is very sensitive to modifications of the form factors. 
 In the SM, the integrated asymmetry is  $\mathscr{A}=15\%$, but it ranges from $-9$\%  to $+27$\% for the illustrative EFT parameters
adopted in Fig.~\ref{fig:spectra}.

\paragraph{\bf V. {\bf Mesonic decays:}}
The $h\to VP$ amplitude depends on the current matrix element
\begin{align}
 \braket{P(q) | J_\mu | 0 }  = \frac12 F_P \, q^\mu~,
\end{align}
where $F_P$ is the pseudoscalar meson decay constant. This current selects a single form-factor combination  
\begin{align}
f^V_P({q}^2) = \frac{ f^V_1(\hat{q}^2) + {q}^2 f^V_2(\hat{q}^2) }{{q}^2- m_V^2}~.
\end{align}
which has no pole as $q^2 \to m_V^2$, by Eq.~(\ref{pole}).
In the SM, one has
\begin{align}
g_V^2 f^{V, \rm SM}_{P}({q}^2)   = - \frac{\gv^2}{m_V^2} = -\frac{4}{v^2}\,, 
\label{eq:f0SM}
\end{align}
which is independent of $\gv$
(here $v = 246 \, {\rm GeV}$).
It is instructive to look at the structure of the amplitude for the $h\to VP$ process,
\begin{align}
\left( \cA^P_V \right)^{\rm SM}  =  -\frac{\gv C_V}{4} \frac{ F_P}{v} (\vareps \cdot  q)   \,. 
\end{align}
This amplitude probes the ratio of  two order parameters which both break
$SU(2)\times U(1)$ in the SM, $F_P$ from the quark condensate 
and $v$ from the Higgs sector. It  provides a very interesting 
probe of the vacuum structure of the theory.\footnote{  The  combination $F_P/v$ appears also in 
the purely leptonic $P\to \ell \nu$ decays; however, in that case it is a probe of the Goldstone component of the 
Yukawa interaction (as manifest from the gauge-less limit of the theory, see e.g.~\cite{FlavLect}). 
 Computing the $h\to VP$ amplitude in the $g\to0$ limit 
treating $V$ as an external field shows that the $h\to VP$ amplitude 
probes the trilinear $h\, \partial_\mu \varphi \, V^\mu$ coupling, where $\varphi$ is a (eaten) Goldstone boson.}

Compared to the leading  decay mode of a light Higgs, $h\to b\bar b$, the $h \to VP$ decay amplitude is parametrically 
suppressed by the small ratio $F_P/m_b$. In the limit where we neglect the mass of the pseudoscalar meson,
\begin{align}
\frac{ \Ga(h\to VP)^{\rm SM } }{  \Ga(h\to b \bar b)^{\rm SM }  }  &=& \frac{m_h^2}{6 v^2} \left| \frac{  C_V F_P }{ \m_b(m_h) } \right|^2 \left(1 -\frac{m_V^2}{m_h^2}\right)^3  
\label{eq:hVPSM}
\end{align}
where  $\Ga(h\to b \bar b)^{\rm SM} = 3 m_h \, \m_b^2(m_h) / (8 \pi v^2)$.
This expression holds both for $V=Z$ and for $V=W^\pm$, separately for each sign of charge.

\begin{table}[t]
\tabcolsep 8pt
\begin{tabular}{|cc|cc|}
\hline
$VP$~mode & $\cB^{\rm SM}$ & $VP^*$~mode & $\cB^{\rm SM}$  \\ \hline
 $W^- \pi^+$  &   $0.6 \times 10^{-5}$  &  $W^- \rho^+ $  &    $0.8 \times 10^{-5}$ \\
 $W^- K^+$  &    $0.4 \times 10^{-6}$  &  $Z^0 \phi$      &    $2.2 \times 10^{-6}$   \\ 
 $Z^0 \pi^0$  &   $0.3 \times 10^{-5}$  &  $Z^0 \rho^0 $  &  $1.2 \times 10^{-6}$  \\
 $W^- D^+_s$  &    $2.1 \times 10^{-5}$ & $W^- D^{*+}_s$  &    $3.5 \times 10^{-5}$ \\
$W^- D^+$      &    $0.7 \times 10^{-6}$   & $W^- D^{*+} $      &    $1.2 \times 10^{-6}$ \\ 
$Z^0 \eta_c$  &   $1.4 \times 10^{-5}$  & $Z^0 J/\psi$  &   $2.2  \times 10^{-6}$\\
\hline
\end{tabular}
\caption{\label{tab:uvbounds} SM branching ratios for selected  $h\to VP$ and $h \to VP^*$ decays. }
\end{table}

Given the normalization of the currents in Eq.~(\ref{currentdefn}), the explicit expressions of $F_P$ in some of the 
most interesting modes are
 $F_{\pi^{\pm}} = V_{ud} f_\pi$, 
$F_{K^\pm} = V_{us} f_K$, $F_{\pi^0} = f_\pi/\sqrt{2}$, $F_{D^\pm}= V_{cd} f_D$, $F_{D_s} = V_{cs} f_{D_s}$,   and $F_{\eta_c} = f_{\eta_c}/2$, 
where $f_P$ are the standard meson decay constants 
$f_\pi \approx 130$~\mev, $f_K \approx 160$~\mev,  $f_D \approx 207$~\mev, 
 $f_{D_s} \approx 250$~\mev, and $f_{\eta_c} \approx 400$~\mev~\cite{Beringer:1900zz,Colangelo:2010et,Davies:2010ip}.
From these values we deduce the SM rates reported in Table~I. Despite the smallness of these rates,
and the huge background at the LHC,
we stress that some channels may have a relatively clean experimental signature, due to the 
displaced vertex of  the subsequent $P$ decay.

Within the  general EFT approach the $h\to VP$ rate assumes the following form 
\begin{align}
\frac{  \Ga(h\to VP)  }{ \Ga(h\to VP)^{\rm SM } } 
&= \abs{c_1 +  g_2^2(c_2 + c_3)}^2\,,  \label{eq:hVPmod}
\end{align}
with possible $\mathcal{O}(1)$ variations with respect to the SM. These variations are closely related to the possible 
variation of the $d\Gamma(h\to V\ell\ell)/d {q}^2$ spectrum at ${q}^2=0$, which is quite difficult to access experimentally.
As a further illustration of the complementarity of $h\to V\ell\ell$ and $h\to VP$ modes, we report here the dependence 
of the two total rates from the EFT parameters, adopting a common normalization for the leading term:
\begin{align}
\Gamma_{V\ell\ell}  &\propto c_1^2 + 0.9\,  c_1  c_2 + 1.3\, c_1 c_3  + 0.6\, c_2 c_3 +0.2\, c_2^2 + 0.5\,  c_3^2~,   \nn
\Gamma_{VP}  &\propto c_1^2 + 0.9\, c_1  c_2 + 0.9\, c_1 \, c_3  + 0.4\, c_2 c_3 +  0.2 \, c_2^2 +  0.2\, c_3^2~.
\end{align}

In the limit where we neglect light hadron masses, Eqs.~(\ref{eq:hVPSM}) and (\ref{eq:hVPmod})
continue to hold with $P \to P^*$, where $P^*$ is a vector meson, with decay constant defined by
\begin{align}
 \braket{P^*(q,\epsilon) | J_\mu | 0 }  = \frac12 F_{P^* }\, m_{P^*}\, \epsilon^\mu~\,.
\end{align}
The corresponding SM rates are given in Table~I using $f_\rho \approx 157$~\mev,  $f_{\phi} \approx 210$~\mev,   $f_{J/\psi} \approx 410$~\mev~\cite{Beringer:1900zz,Donald:2012ga} and $f_{D^*_{(s)}}/f_{D_{(s)}} \approx 1.3$~\cite{Abada:1991mt}. The $P^*$ is longitudinally polarized. For heavy quarks, spin symmetry implies $f_P=f_{P^*}$ \cite{MW}; the vector structure of  $J^Z_\mu$
implies $F_{\rho^0} = (1-2s_W^2) f_\rho/\sqrt{2}$, $F_{\phi} = (1/2-2/3 s_W^2) f_\phi$, 
$F_{J/\psi} = (1/2-4/3 s_W^2) f_{J/\psi}$.

{\bf VI. {\bf Conclusions:}}\label{conclusions}
We have shown the importance and utility of a decomposition of the $h \rightarrow V \fstate$  amplitude into form-factors which can be probed by different final states, and how differential spectra can be used to disentangle the $h V  V^*$ couplings  in a general EFT approach. Complementary information is provided by leptonic and exclusive (semi-)hadronic final states, among which the $h \to V P$ decay is a simple and particularly interesting example. 
See~\cite{inprep} for a related analysis in the associated production process.

{\bf Acknowledgements} 
A.M.~was supported in part by DOE Grant No. DE-FG02-90ER40546.
G.I.~was supported in part by  MIUR
under project 2010YJ2NYW.

\vspace{-0.7cm}

\end{document}